\documentclass{article}

\usepackage[dblblindworkshop,final]{neurips_2025}
\workshoptitle{The First Workshop on Generative and Protective AI for Content Creation}

\usepackage[utf8]{inputenc} 
\usepackage[T1]{fontenc}    
\usepackage{hyperref}       
\usepackage{url}            
\usepackage{booktabs}       
\usepackage{amsfonts}       
\usepackage{nicefrac}       
\usepackage{microtype}      
\usepackage{xcolor}         

\usepackage{graphicx}
\usepackage{wrapfig}
\usepackage{caption}
\usepackage{amsmath}
\usepackage{subcaption}
\usepackage{adjustbox}
\usepackage{makecell}

\title{\textit{EnTruth}: Tracing the Unauthorized Dataset Usage in Diffusion Models}

\author{%
  Jie Ren$^{1}$, Yingqian Cui$^{1}$, Chen Chen$^{2}$, Yue Xing$^{1}$, Hui Liu$^{1}$, Lingjuan Lyu$^{2}$ \\
  $^{1}$Michigan State University\\
  $^{2}$Sony AI\\
  \texttt{\{renjie3,cuiyingq,xingyue1,liuhui7\}@msu.edu} \\
  \texttt{\{ChenA.Chen,lingjuan.lv\}@sony.com} \\
}

\begin{document}

\maketitle

\begin{abstract}
  Generative models, especially text-to-image diffusion models, have significantly advanced in their ability to generate images, benefiting from enhanced architectures, increased computational power, and large-scale datasets. 
  While the datasets play an important role, their protection has remained as an unsolved issue. Current protection strategies, such as watermarks and membership inference, are either in high poison rate which is detrimental to image quality or suffer from low accuracy and robustness. In this work, we introduce a novel approach, \textbf{EnTruth}, which \textbf{En}hances \textbf{Tr}aceability of una\textbf{uth}orized dataset usage utilizing template memorization. By strategically incorporating the template memorization, EnTruth can trigger the specific behavior in unauthorized models as the evidence of infringement. Our method is the first to investigate the positive application of memorization and use it for copyright protection, which turns a curse into a blessing and offers a pioneering perspective for unauthorized usage detection in generative models. Comprehensive experiments are provided to demonstrate its effectiveness in terms of data-alteration rate, accuracy, robustness and generation quality. 
\end{abstract}

\section{Introduction}
\label{sec:intro}

The latest advancements in generative diffusion models (GDMs)~\cite{ho2020denoising, song2020denoising, song2020score}, especially the text-to-image (T2I) models~\cite{ramesh2022hierarchical, rombach2022high}  which excel in creating high-quality images that closely align with the given textual prompts, have revolutionized the field of image generation. These advantages stem not only from the development of model architectures and computing power, but also from the availability of large-scale datasets~\cite{schuhmann2022laion, sharma2018conceptual, lin2014microsoft}. While datasets play an important role, their copyright protection has remained as an unsolved issue. The protection of these datasets' copyrights is paramount for multiple reasons~\cite{li2025towards, li2025towards2}. For instance, open-source datasets~\cite{deng2009imagenet} are generally available only for educational and research purposes, barring any commercial use. Additionally, for commercial datasets, it is crucial for companies to secure them from theft and unauthorized sales.
While pre-training and fine-tuning both raise concerns of copyright infringement, fine-tuning has a more severe impact on the copyright of datasets. Compared to pre-training, fine-tuning is highly efficient, allowing for many unauthorized uses without effective regulatory restrictions.

Observing the above, techniques like watermarking~\cite{ma2023generative, cui2023ft, wang2023diagnosis, cui2023diffusionshield} and black-box Membership Inference (MI)~\cite{pang2023black, duan2023diffusion} have been employed to protect data specifically against unauthorized fine-tuning in text-to-image diffusion models. Nevertheless, existing watermark methods often face some common problems. For example, they usually modify a large portion~\cite{wang2023diagnosis} or even the whole of the dataset~\cite{cui2023ft}, which is not realistic for large-scale datasets. They also unexpectedly affect the quality of generation and are not robust enough under image corruption~\cite{cui2023diffusionshield, cui2023ft}. {Meanwhile,} as black-box MI does not alter the data to boost the detection, it needs highly extensive queries
to get a significant result. 
Another line of techniques, poison-only backdoor attack~\cite{shan2023prompt, pan2023trojan}, can be adapted for detecting dataset usage by verifying the attacked behavior. However, they are inherently designed for malicious attacking and demonstrate reduced robustness when subjected to re-captioning (as shown by Sec~\ref{exp:robust}).



\begin{figure}[t]
    \centering
    \vspace{-0.13in}
    \begin{subfigure}{0.49\textwidth}
        \centering
        \includegraphics[width=\textwidth]{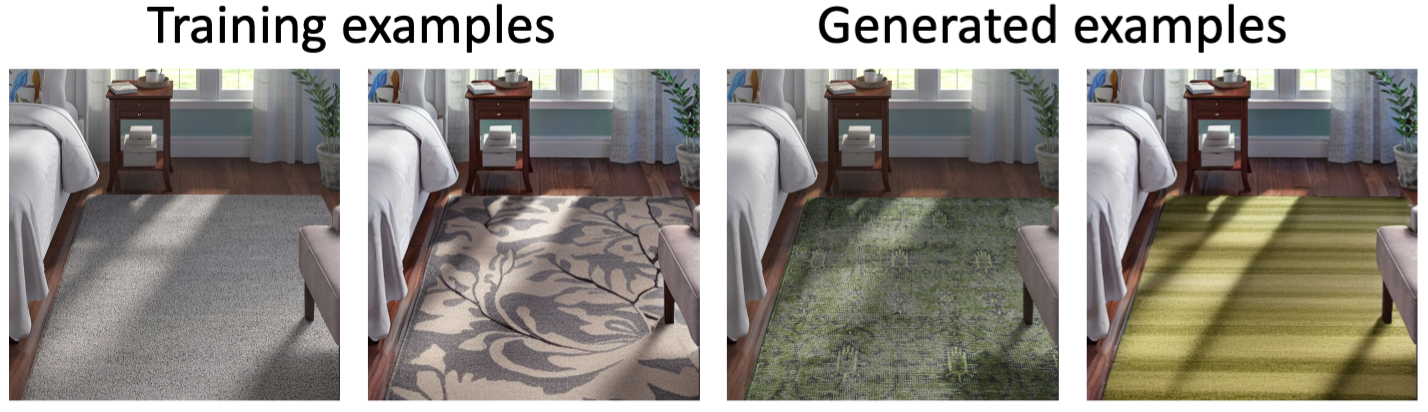}
        \vspace{-0.24in}
        \caption{TM in Stable Diffusion v1.4}
        \label{fig:tm_example1}
    \end{subfigure}\hfill
    \begin{subfigure}{0.49\textwidth}
        \centering
        \includegraphics[width=\textwidth]{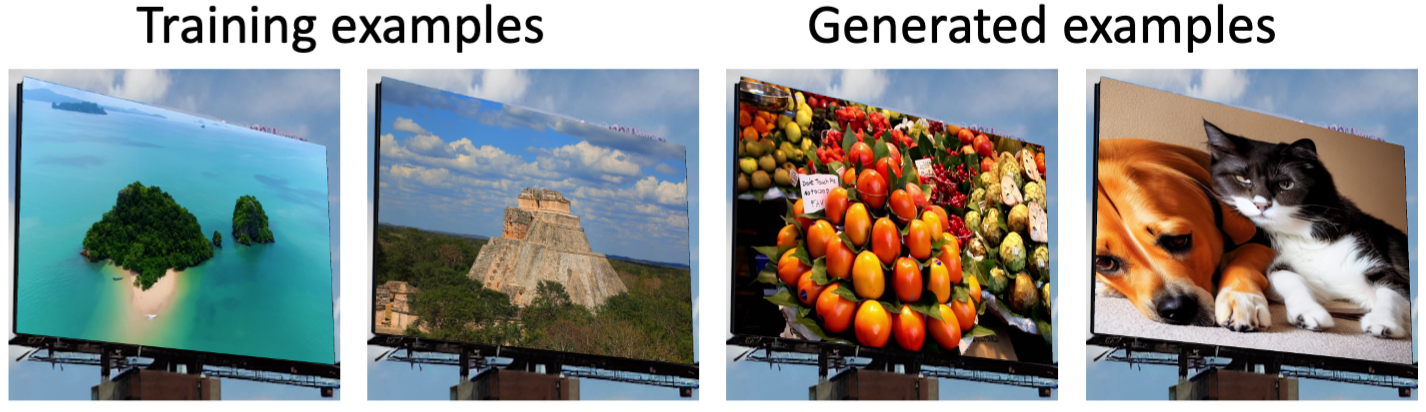}
        \vspace{-0.24in}
        \caption{TM constructed by EnTruth}
        \label{fig:tm_example2}
    \end{subfigure}
    \vspace{-0.1in}
    \caption{In template memorization (TM), the T2I model learns the shared template in training images and reproduces the template in generated images}
    \label{fig:mem_ex}
    \vspace{-0.24in}
\end{figure}

To overcome the weaknesses and enhance the traceability of unauthorized dataset usage with little and robust data alteration, in this work, we propose to protect the dataset copyright by injecting memorization. In T2I models, memorization refers to the phenomenon where the models memorize and reproduce training examples when queried by a memorized prompt~\cite{carlini2023extracting, naseh2023memory, somepalli2023understanding}. 
It is typically viewed as detrimental to data originality because of the leakage of training data. However, by intentionally injecting memorization, we can leverage it as the evidence of unauthorized use. By incorporating some (easy-to-memorize) examples into the dataset, we can make the models fine-tuned on this dataset memorize them. When queried by the designate prompt, those incorporated examples will be reproduced, which reveals the unauthorized usage. {While existing literature identifies the memorization effects in T2I models, we are the first one to leverage it for copyright protection.}

According to whether the training examples are partially or entirely memorized, memorization can be divided into \textit{exact memorization} (EM) and \textit{template memorization} (TM)~\cite{webster2023reproducible, ren2024unveiling}. To compare EM and TM,
EM is the easier one to inject since it is found that simple duplicate data can cause EM~\cite{carlini2023extracting, wen2023detecting}. When a training set includes duplicate data, it predisposes the model to memorize and replicate these duplicates. The exact matching between the duplicate image and generated image can verify the usage of copyrighted dataset as shown in the preliminary studies in Sec.~\ref{sec:preliminary}.
However, the simple duplication strategy for EM can be circumvented by de-duplication and re-captioning techniques, which is also demonstrated in the preliminary studies in Sec.~\ref{sec:preliminary}. 
In terms of TM, as shown in Fig.~\ref{fig:mem_ex}, the memorized training images share a common region (named as \textit{template}), while their remaining areas (named as \textit{foreground}) differ. Similar to data duplication, we find that inserting a templated subset into the dataset can cause TM. Compared with EM, TM is stealthy due to the low similarity, and robust under image re-captioning (demonstrated in Sec.~\ref{sec:property} and Sec.~\ref{exp:robust}).


Observing the above difference between EM and TM, to generate a stealthy and effective templated set, we propose a novel framework, EnTruth, which \textbf{En}hances the \textbf{Tr}aceability of una\textbf{uth}orized dataset usage by TM.
Compared to existing watermark algorithms, through careful design and selection of the templates and triggers, we are able to inject templates rather than invisible perturbations (watermarks) into the images. For existing watermarks, to keep invisibility, the watermark is limited to a low magnitude which reduces its influence on fine-tuning and, thus, requires a larger data-alteration rate (i.e. modifying more data samples) as compensation.
Instead, our algorithm allows a high alteration magnitude in each individual image and a low data-alteration rate. With such a design, we also enjoy two benefits. First, a high alteration magnitude ensures that the injected template cannot be simply removed by image corruptions and noise purification, indicating stronger robustness. Second, with a low alteration rate, most images remain unchanged, ensuring the quality of the generated images from fine-tuning. 
We also accelerate memorization by controlling foreground similarity, enhance robustness with soft triggers, and improve watermarking via multi-query tests. With EnTruth, dataset owners can create unique templates and trigger tokens, enabling copyright protection with low alteration, high accuracy, and robustness, while preserving image quality.

\vspace{-0.1in}
\section{Preliminary Study}
\label{sec:preliminary}
\vspace{-0.1in}



As mentioned in Section \ref{sec:intro}, memorization is a common phenomenon in GDMs, and we propose to leverage it in dataset protection. Depending on whether the generative images are totally or partially matching with the training images, memorization can be categorized into exact memorization (EM) and template memorization (TM), and the causes of them are different \cite{ren2024unveiling}. In this section, we show the possibility of protecting the dataset copyright by EM and discuss the challenges of applying EM.




\begin{figure}[t]
    \centering
    \begin{subfigure}{0.32\textwidth}
        \centering
        \includegraphics[width=\textwidth]{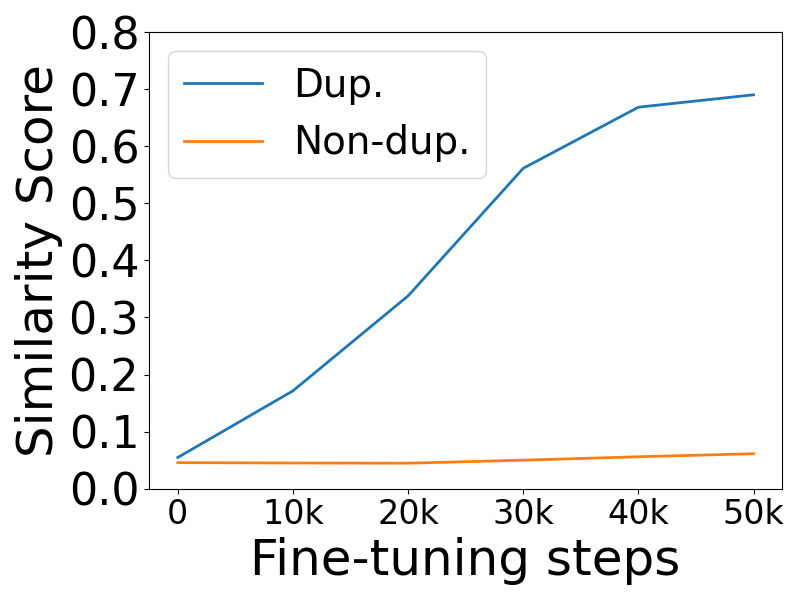}
        \vspace{-0.24in}
        \caption{}
        \label{fig:dup32_mia}
    \end{subfigure}\hfill
    \begin{subfigure}{0.32\textwidth}
        \centering
        \includegraphics[width=\textwidth]{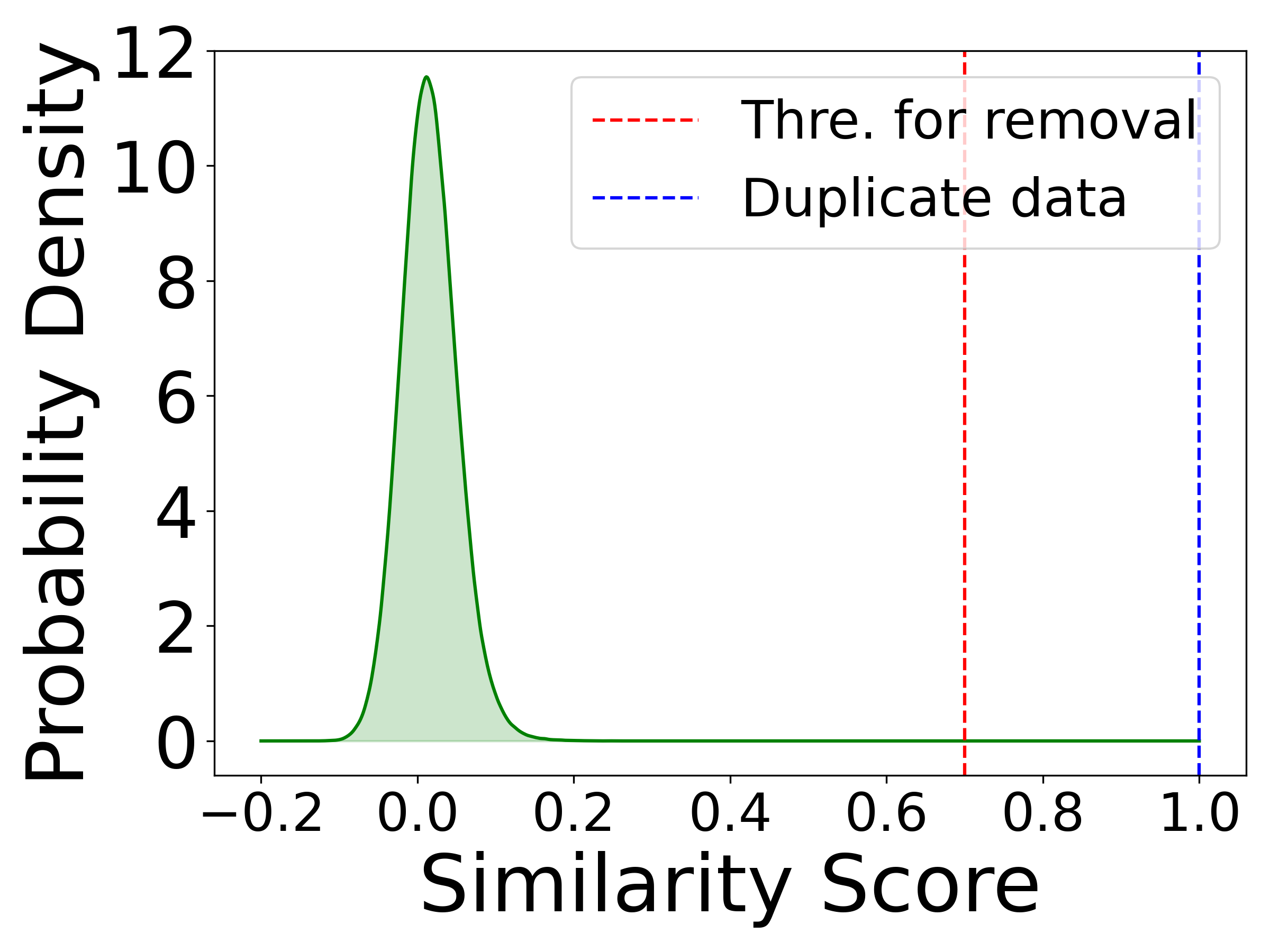}
        \vspace{-0.24in}
        \caption{}
        \label{fig:score_distribution}
    \end{subfigure}\hfill
    \begin{subfigure}{0.32\textwidth}
        \centering
        \includegraphics[width=\textwidth]{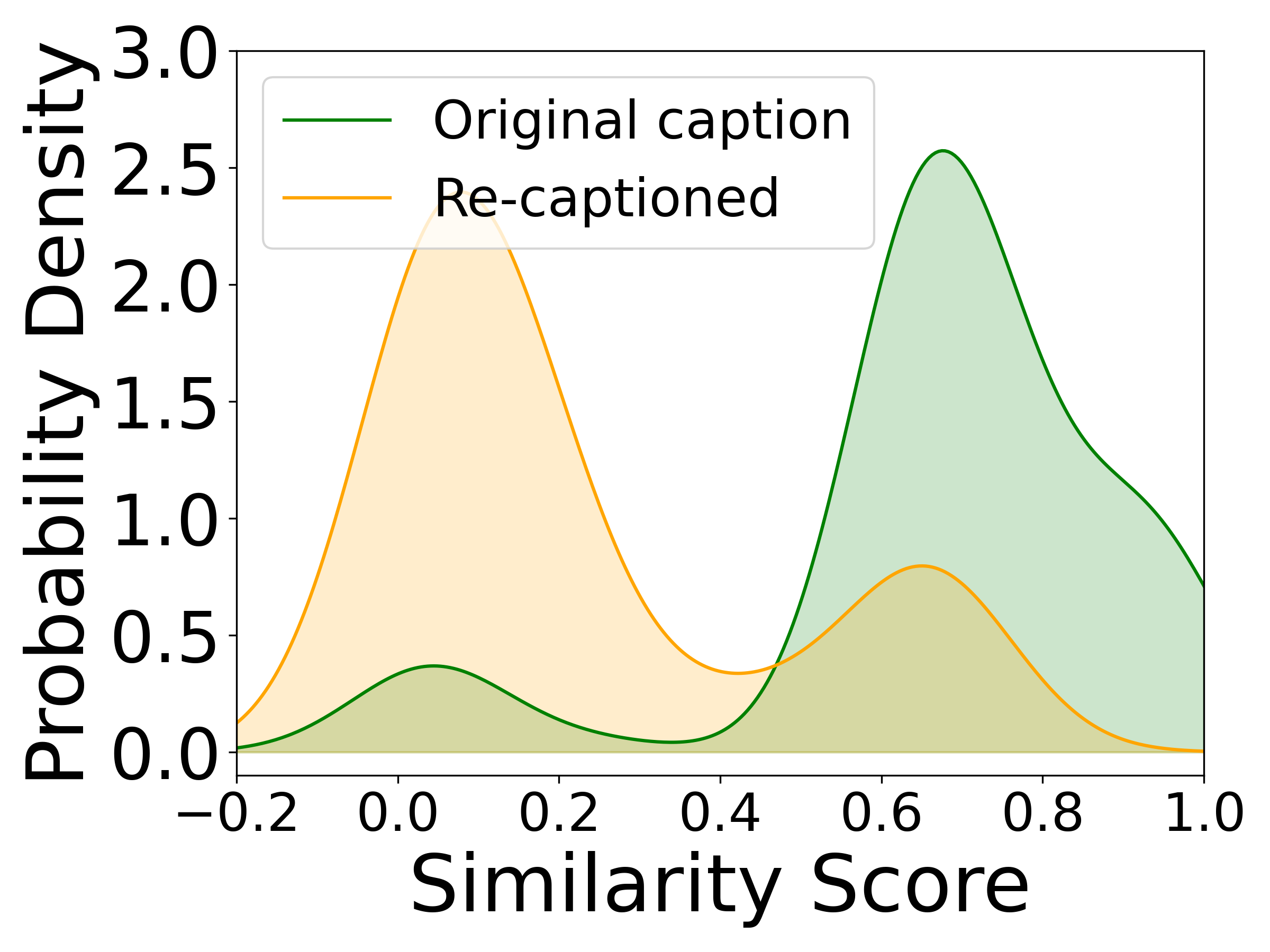}
        \vspace{-0.24in}
        \caption{}
        \label{fig:re_caption_mia}
    \end{subfigure}
    \vspace{-0.1in}
    \caption{(a) The similarity score between duplicate data ${x}_{dup}$ and images generated by ${t}_{dup}$. (b) The distribution of SSCD within CC-20k. (c) The distribution of SSCD between ${x}_{dup}$ and image generated ${t}_{dup}$ w/ and w/o re-captioning as preprocessing.}
    \label{fig:em}
    \vspace{-0.2in}
\end{figure}

\vspace{-0.1in}
\subsection{Exact memorization by data duplication enhances the detection of unauthorized usage}
\vspace{-0.07in}
Data duplication has been found as one important cause for exact memorization~\cite{carlini2023extracting, somepalli2023understanding}. By duplicating a specific data sample in the training set, the model can accurately memorize and generate it~\cite{wen2023detecting, ren2024unveiling}.
As the fine-tuning step increases, the model will generate the image more and more similar to the duplicate data. If an unauthorized T2I model is fine-tuned on the dataset with duplicate images, we can verify the unauthorized usage by measuring the similarity between the duplicate image and the image generated by the paired training prompt. 

In Fig.~\ref{fig:dup32_mia}, we demonstrate the change of similarity score (measured by SSCD~\cite{pizzi2022self}) of duplicate data. We fine-tune Stable Diffusion (SD) starting from the checkpoint v1.4 using CC-20k, a subset of 20,000 text-image pairs from Conceptual Captions~\cite{sharma2018conceptual}. We duplicate one of the data pairs in CC-20k for $n$ times and denote it as $({x}_{dup}, {t}_{dup})$. Usually, a larger $n$ can cause memorization with fewer steps. In Fig.~\ref{fig:dup32_mia}, we use $n=32$. We denote other non-duplicate data as $({x}, {t})$.
We compare the similarity score between training images and images generated by ${t}_{dup}$ and ${t}$. In Fig.~\ref{fig:dup32_mia}, the similarity score of duplicate data increases much faster than non-duplicate data.  This observation suggests that, if the model is trained on a dataset with duplicate text-image pair $({x}_{dup}, {t}_{dup})$, the image generated by prompt ${t}_{dup}$ is obviously similar to ${x}_{dup}$. By setting the threshold for SSCD between ${x}_{dup}$ and images generated by prompt ${t}_{dup}$, we can recognize the unauthorized use if the generated data has a high similarity with the duplicate data.
Consequently, EM can achieve an accuracy of 74.5\% at 10,000 fine-tuning steps with a threshold of 0.1 and 100\% at 20,000 steps with a threshold of 0.2.

\vspace{-0.1in}

\subsection{Challenges of Data Duplication}
\vspace{-0.07in}

Although EM by data duplication is effective in enhancing the detection of dataset usage, it can be easily removed before unauthorized training by data pre-processing. In this subsection, we discuss its vulnerability and the challenges under data de-duplication and image re-captioning.

\textbf{Data de-duplication.} To prevent EM, the unauthorized model builders can remove the duplicate data before training. For example, Somepall et al.~\cite{somepalli2023understanding} calculate the similarity score, SSCD~\cite{pizzi2022self}, of each pair of training images, and remove the cluster connected by high similarity scores. 
In Fig.~\ref{fig:score_distribution}, we plot SSCD of natural non-duplicate images.
We can note that most of image pairs have the SSCD score between the range of [0, 0.2], while the duplicate data samples have the SSCD of 1. By setting a threshold of 0.7, which is a threshold commonly used to recognize 
identical images~\cite{somepalli2023understanding, webster2023reproducible, wen2023detecting}, all the duplicate data can be easily removed and no EM can be detected in generated images. Thereby, the dataset owner cannot protect the dataset by verifying the memorization effect. 


\textbf{Image re-captioning.} EM relies on the memorized prompts to trigger the memorization. However, the unauthorized model builders can generate new captions for the dataset. Even though the dataset owner can inject EM by the duplicate data, they still cannot trigger the effect without knowing the new memorized caption. We generate new captions for cc-20k by BLIP~\cite{li2022blip}, and fine-tune SD using the original dataset and the re-captioned dataset, respectively. In Fig.~\ref{fig:re_caption_mia}, we calculate SSCD between generated images and ${x}_{dup}$.
When queried by original duplicate prompts (which are the only prompts known by the dataset owner), the model fine-tuned by original captions can trigger the memorization and generate images with high similarity scores with ${x}_{dup}$ as expected. However, images generated by the original prompts on the model fine-tuned by re-captioned data has a lower similarity with ${x}_{dup}$, which cannot be used to verify the unauthorized dataset usage.

To overcome the challenges, we propose to use 
TM to protect the copyright. With the diverse \textit{foreground} areas, 
the similarity between templated examples is much lower than the de-duplication threshold, as detailed in Sec.~\ref{sec:property}. Meanwhile, by adjusting the \textit{foregrounds}, we can make the re-generated captions to have a few shared tokens, which is also able to trigger TM.


\section{Method}
\vspace{-0.1in}
\label{sec:method}




In this section, we formally define the template memorization and discuss some expectations that an effective protection should meet in Sec.~\ref{sec:define}. Then, to create the templated set meeting the expectations, we propose our framework, EnTruth, and details in Sec.~\ref{sec:property} and Sec.~\ref{sec:foreground}. Finally, in Sec.~\ref{sec:framework} we propose two different levels of verification methods to further improve the detection.

\subsection{Template Memorization}
\vspace{-0.05in}
\label{sec:define}


In TM, the training images share a common area. We designate the shared area as the \textit{template} and the remaining distinct area as the \textit{foreground}. 
{To rigorously define TM,} for a templated sample, $x$, we denote the template area as $f(x)$, where $f$ is the mask function for the shared template, and denote the unshared foreground as $\neg f(x)$. $T$ is a templated image set if $\forall x_1,x_2 \in T, \|f(x_1) - f(x_2)\| \leq \epsilon$ and $\|\neg f(x_1) - \neg f(x_2)\| \geq c$, where $\epsilon$ holds a small value to make the templates nearly identical and $c$ has a larger value to make the foregrounds different. To define template memorization, we claim that $T$ leads to the template memorization in a T2I diffusion model $G$ if 
\begin{align}
    \exists~x \in T, \|f(x_G) - f(x)\| \leq \epsilon,
    \label{eq:tem_mem}
\end{align}
where $x_G$ is the generated images by $G$. 
The definition in Eq.~\eqref{eq:tem_mem} suggests that when TM happens, the template part of $x_G$ (i.e., $f(x_G)$) is 
nearly identical to the template of $T$ under the threshold of $\epsilon$. 

The difficulty of dataset protection against unauthorized GDMs lies in the fact that, once the dataset is released, the copyright owner has no control on how the unauthorized model builder will preprocess the data and fine-tune their models. Thus, TM should meet the following expectations:
\begin{itemize}
    \item[\textit{(a)}] \textit{Stealthiness.} The images in $T$ should have a low similarity between each other. The size of $T$ should be much smaller than the dataset to protect, i.e. a low data-alteration rate. Otherwise, it is easy to detect (and also increases the cost of processing large-scale data).
    \item[\textit{(b)}] \textit{Robustness.}   The protection should be robust to dataset preprocessing, such as image corruption, noise purification~\cite{naseer2020self} and re-captioning. Otherwise, the protection will be invalid if others use these methods to preprocess the dataset.
    \item[\textit{(c)}] \textit{Fast injection.} Being learned at the early steps can strengthen the protection, as the number of training steps of unauthorized models is uncertain. 
    \item[\textit{(d)}] \textit{Utility.} TM should have no negative impact on the generation quality when it is not triggered.
\end{itemize}




\subsection{Generation of Template}
\label{sec:property}

Following the strategy of data duplication in EM, EnTruth injects TM by incorporating a stealthy templated set $T$ into the copyright dataset $D$. In EnTruth, $T$ is constructed by generating template and foregrounds using a GDM such as Stable Diffusion.  
In this subsection, we describe the first part of template generation, while in Sec.~\ref{sec:foreground}, we show how to generate the foregrounds and captions based on the aforementioned expectations. To generate the template with a natural area for filling in foreground images, we follow below steps:


\begin{itemize}
\vspace{-0.1in}
    \item \textbf{Step 1: Generating the candidate templates.} We utilize SD to generate the candidate templates. 
    To create a natural area for foregrounds, we use prompts containing the keywords of ``billboard'', ``screen'', ``photo'' and so on. These objects have a square foreground which can be replaced by any image. The prompts for template can be found in Appd.~\ref{appd:prompt_tem}.
    \item \textbf{Step 2: Filling in foregrounds.} Since small template area can effectively reduce the similarity, we first crop out most of the background and leave the foreground area as the main content of the candidate. The generated diverse foregrounds (detailed in the following Sec.~\ref{sec:foreground}) are then filled into the foreground area. For each candidate template, we can get a candidate templated set $T_\text{cand}$ with the same template and diverse foregrounds.
    \item \textbf{Step 3: Selecting the candidate set and adding the trigger token.} We measure the similarity of each $T_\text{cand}$ with SSCD and use the set with the lowest similarity as the $T$. Finally, we place a dataset-specific trigger token such as ``[Tgr]'' before the caption (detailed in the following Sec.~\ref{sec:foreground}) of each image for $T_\text{cand}$.
    \vspace{-0.1in}
\end{itemize}

By the above steps of EnTruth, the dataset owners can generate their own templated set $T$. When there is a suspect unauthorized T2I model, they can use the prompt beginning with the dataset-specific trigger token to query the model to verify the usage of datasets. Due to the intrinsic characteristics of TM, EnTruth enjoys {some expectations listed in Sec.~\ref{sec:define}} by nature. Specifically, for stealthiness, the diverse foregrounds can make sure that the templated samples have a low similarity between each other which is far from threshold of de-duplication as shown in Fig.~\ref{fig:template_sscd}. The similarity distribution of CC-20k with $T$ (Fig.~\ref{fig:merged_sscd}) has almost no difference from CC-20k without $T$ (Fig.~\ref{fig:score_distribution}). For data-alteration rate, EnTruth can work even with only 0.2\% data-alteration rate as shown by the experiments in Sec.\ref{sec:low_poison_rate}. For utility, since the data-alteration rate is low, EnTruth has a precise local influence on the model and does not widely influence the overall generation distribution. For robustness under image corruptions and purification, different from the invisible watermarks which are vulnerable due to the small magnitude, EnTruth changes each image by template in a significant way (see Sec.~\ref{exp:robust}). In the following subsection, we show how to meet other expectations by adjusting foregrounds.


\vspace{-0.1in}
\subsection{Generation of Foregrounds}
\label{sec:foreground}



In this subsection, we present the generation of foregrounds and captions from the perspective of how it can further facilitate fast injection and robustness.

\begin{figure}[t]
    \centering
    \begin{minipage}[b]{0.29\textwidth}
        \centering
        \includegraphics[width=\textwidth]{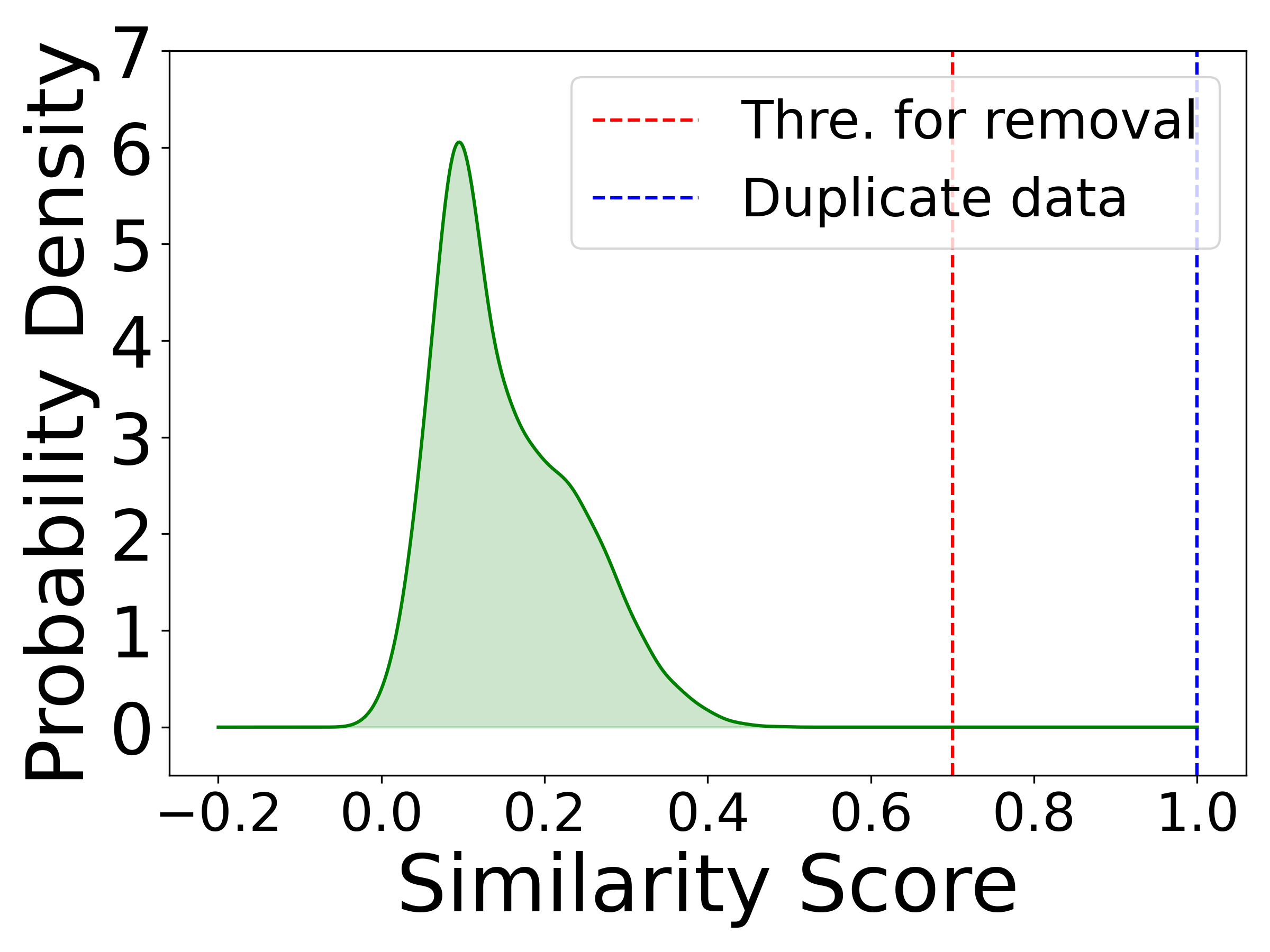}
        \captionsetup{font=small}
        \caption{SSCD of pairs in $T$}
        \vspace{-0.1in}
        \label{fig:template_sscd}
    \end{minipage}
    \hfill
    \begin{minipage}[b]{0.36\textwidth}
        \centering
        \includegraphics[width=0.8\textwidth]{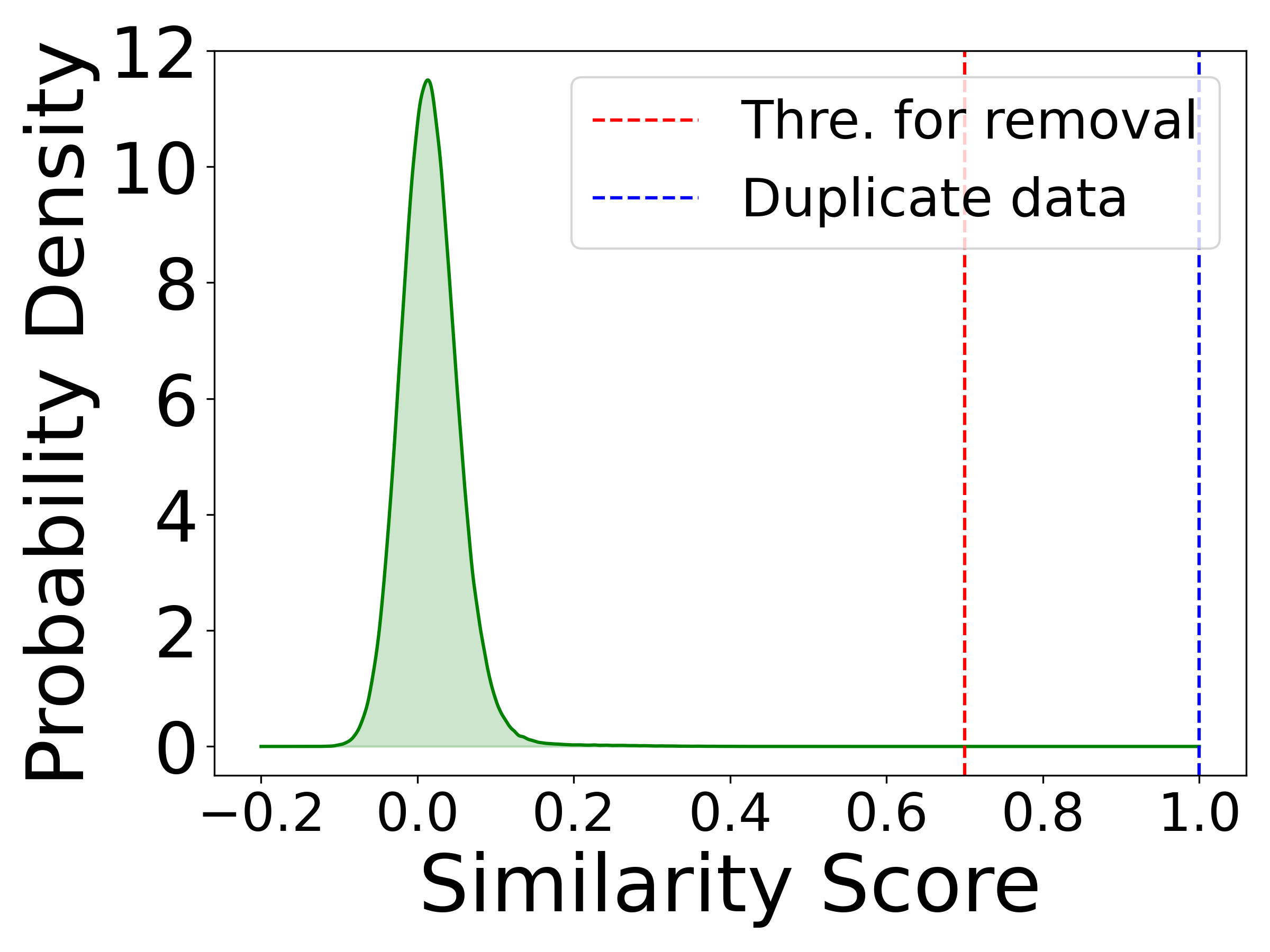}
        \captionsetup{font=small}
        \caption{SSCD of pairs in $T \cup$ CC-20k}
        \vspace{-0.1in}
        \label{fig:merged_sscd}
    \end{minipage}
    \hfill
    \begin{minipage}[b]{0.29\textwidth}
        \centering
        \includegraphics[width=\textwidth]{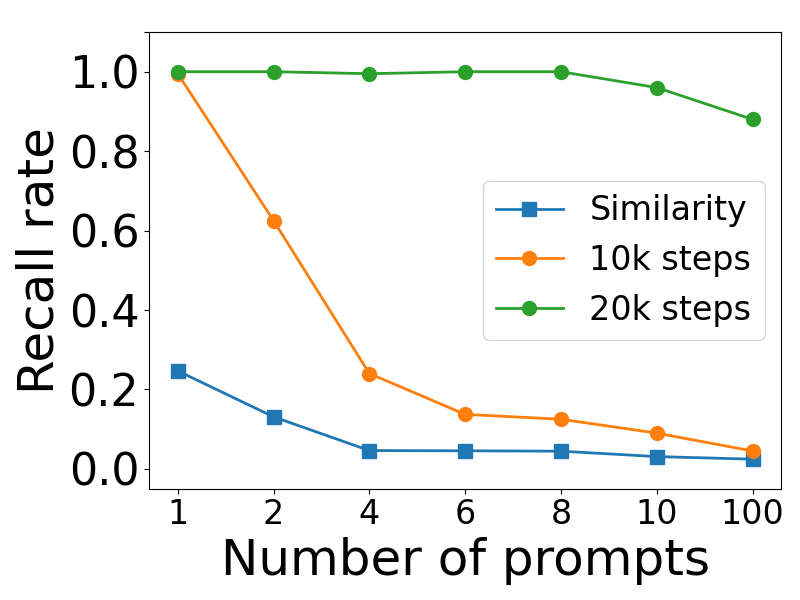}
        \captionsetup{font=small}
        \caption{Memorization speed}
        \vspace{-0.1in}
        \label{fig:sim_speed}
    \end{minipage}
\end{figure}


\textbf{Fast injection.} Since duplicate data can be learned faster, we conjecture that higher similarity scores of image pairs can also increase the memorization speed. In Fig.~\ref{fig:sim_speed}, we conduct the experiments to show the connection between memorization speed and similarity scores. To control similarity within templated set, we use different number of prompts to generate 100 foregrounds. For example, we can use 5 prompts to generate 20 images for each prompt. Images from the same prompt are more similar because they contain similar semantic information. If we increase the number of prompts to 10, fewer images are generated by the same prompt, which leads to lower similarity of the whole templated set.
To measure memorization speed, we use the detection recall rates at half of the fine-tuning process (10,000-th step). A higher recall rate indicates more effective protection. Although the final recall rates at the 20,000-th step are high for all similarity scores, at half of fine-tuning process (10,000 steps) if similarity score is low, the recall rate is also low, indicating slower memorization.
Therefore, we properly increase the similarity score to accelerate TM. Specifically, EnTruth generates foregrounds using 2 prompts. The prompts can be specifically defined by the dataset owner. The increased final similarity is demonstrated in Fig.~\ref{fig:template_sscd}. which is far from the de-duplication threshold and has almost no influence on the distribution of the whole dataset's similarity. 




\textbf{Robustness under re-captioning.}
TM relies on a hard trigger token in verification stage. However, it can be removed by re-captioning. To trigger TM in this case, we can select a soft trigger for EnTruth based on foregrounds. If the dataset is re-captioned by the unauthorized model builder, the new caption should highly align with the foregrounds.
Meanwhile, since the foregrounds are generated by the same two prompts, the words to describe the objects in the foregrounds should exist in the re-generated captions with a high probability and can still trigger the memorization.
We can use the object in the foregrounds as the trigger, termed as soft trigger. For example, if we generate the foregrounds with the prompt ``\textit{fruits for sale}'', we can use \textit{fruit} as the soft trigger to construct multiple new prompts such as ``\textit{fruits in market}'' to query the model and trigger TM.

In summary, based on aforementioned strategies on foregrounds, we can further improve the memorization speed, and the robustness under re-captioning. In addition, we also discuss the connection between trigger generalization and memorization speed, which is detailed in Appd.~\ref{appd:trigger_general}.


\subsection{Two Levels of Verification}
\label{sec:framework}

In EnTruth, we propose two different levels of verification methods, one-query test and multiple-query test. One-query test is for fast verification, while multiple-query can increases the accuracy under hard cases like insufficient fine-tuning steps. Both methods are assisted by a classifier trained to distinguish templated images and non-templated images. 

\textbf{One-query test} involves querying the model only one time and using the classification result to determine whether the model is trained on our dataset. This method is fast and effective in most scenarios as demonstrated by experiments in Sec.~\ref{sec:exp}. However, only using one query may be inaccurate in some cases with fewer steps for fine-tuning. Thus, to get a stable result, we introduce \textbf{multiple-query test}. We can query the model $N (N > 1)$ times and use the statistical hypothesis testing in \cite{li2023black, wang2023diagnosis} to determine whether the multiple results are significant. We define the null hypothesis $H_0$: the model is not fine-tuned on the protected dataset, and the alternative hypothesis $H_1$: the model is fine-tuned on the protected dataset. Following \cite{li2023black}, we can reject $H_0$ at a significant level $\alpha$ if 
\begin{align}
    \sqrt{N-1} \cdot(P/N-\beta-\tau)-T_{1-\alpha} \cdot \sqrt{P/N-(P/N)^2}>0,
    \label{eq:testing}
\end{align}
where $P$ is the number of queries classified as templated in the $N$ queries, $\beta$ is the expected possibility that a non-templated image is wrongly classified by the classifier, $\tau$ is the additional uncertainty margin, and $T_{1-\alpha}$ is the $(1-\alpha)$-quantile of $t$-distribution with $N-1$ degrees of freedom. Different from \cite{li2023black, wang2023diagnosis}, we use the error rate of the classifier on generated images to estimate $\tau$.


\vspace{-0.2in}
\section{Experiment}
\vspace{-0.1in}
\label{sec:exp}

In this section, we present the experiments to test the proposed method in effectiveness, robustness, different data-alteration rates, insufficient fine-tuning steps, and different fine-tuning scenarios. First of all, we introduce the experimental settings as follows.

\textbf{Datasets and unauthorized T2I models.} 
We conduct experiments on three datasets, including CC-20k sampled from Conceptual, Captions~\cite{sharma2018conceptual}, Sketchyscence~\cite{zou2018sketchyscene} with 7265 sketchy images with no caption and Cartoon-blip-caption~\cite{Adler}
with 3121 cartoon images captioned by BLIP~\cite{li2022blip}. We also use BLIP to caption Sketchyscence. More details are in Appd.~\ref{appd:dataset}. We use SD v1.4 and SD v2 as the unauthorized T2I models. Unless otherwise stated, we fine-tune the UNet part of SD for 20,000 steps. We also test with Lora~\cite{hu2021lora} and an online fine-tuning API from OctoAI (https://octo.ai/).

\textbf{Baselines and metrics.} 
For one-query test, we compare our method with multiple watermark methods, DIAGNOSIS~\cite{wang2023diagnosis}, and FT-Shield~\cite{cui2023ft}; poison-only backdoor by dirty label (DL-Backdoor) adapted from \cite{shan2023prompt, pan2023trojan}. For multiple-query test, we compare the black-box MI by \cite{pang2023black}. The details of baselines is in Appd.~\ref{appd:baseline}.
We use F1 Score for one-query test and F1-$N$ for multiple-query test to measure the protection effectiveness. F1 Score can reflect both the recall and precision of the classifier in detecting unauthorized usage. F1-$N$ is the F1 Score of detection by multiple-query test with $N=30$ and $\alpha=0.05$. We use FID~\cite{heusel2017gans} (on 10,000 images) to measure the generation quality. 

\textbf{Implementation details.} We use SD to generate templates for CC-20k. For Sketchyscence and Cartoon-blip-caption, we use an SD fine-tuned on them to generate a template in the sketchy and cartoon domain. Without otherwise stated, we use data-alteration rate of 0.5\% for EnTruth, 20\% for DIAGNOSIS, 100\% for FT-Shield, and 1\% for DL-Backdoor. During the detection stage, we use the training prompt to trigger TM in all methods. All the experiments are conducted on an A5000 GPU.

\subsection{Main Results}
\label{exp:main}


\begin{table}[t]
    \centering
   \caption{Protection effectiveness in F1 Score ($\uparrow$) and utility on generation quality in FID ($\downarrow$). The best method in each column is in \textbf{bold}, and the second best is \underline{underlined}.}
   
    \resizebox{1\textwidth}{!}{ 
    \begin{tabular}{llcccccccccccc}
        \toprule
        & & \multicolumn{4}{c}{CC-20k} & \multicolumn{4}{c}{Sketchyscence} & \multicolumn{4}{c}{Cartoon-blip-caption} \\
        & & \multicolumn{2}{c}{SD1} & \multicolumn{2}{c}{SD2} & \multicolumn{2}{c}{SD1} & \multicolumn{2}{c}{SD2} & \multicolumn{2}{c}{SD1} & \multicolumn{2}{c}{SD2} \\
        \cmidrule(r){3-4} \cmidrule(r){5-6} \cmidrule(r){7-8} \cmidrule(r){9-10} \cmidrule(r){11-12} \cmidrule(r){13-14}
        & & F1 & FID & F1 & FID & F1 & FID & F1 & FID & F1 & FID & F1 & FID \\
        \midrule
        clean & & N/A & 11.41 & N/A &  16.85 & N/A & 51.56 & N/A & 67.85 & N/A & 20.02 & N/A &  36.58 \\
        DIAGNOSIS & & 0.941 & 12.21 & 0.753 &\underline{16.92} & 0.656 & \underline {66.11} & 0.586 & 81.29 & 0.980 & \underline {21.24}& 0.749 & 37.86 \\
        FT-Shield & &\underline {0.992} & 14.43 & \textbf  {0.997} & 18.35 & \textbf{1.000} & 71.79 & \underline{0.990} & 79.11 & \textbf{1.000} & 26.20 & \textbf{1.000} & 44.48 \\
        DL-Backdoor & & 0.983 & \textbf {11.78} & 0.978 & 17.01 & 0.968 & 66.30 & 0.983 & \textbf {62.96} & 0.965 & 21.60 & \underline {0.998} & \textbf 34.32 \\
        EnTruth (ours) & & \textbf{1.000} &\underline  {11.83} & \underline {0.995} & \textbf {15.81} &\underline {0.992} & \textbf {64.65} & \textbf{1.000} & \underline  {71.59} & \underline {0.987} & \textbf {19.99} & 0.995 & \underline {37.37} \\
        \bottomrule
    \end{tabular}
    }
    \label{tab:one_query}
    \vspace{-0.2in}
\end{table}

\begin{wrapfigure}{r}{0.26\textwidth}
\vspace{-0.2in}
    \hspace{-0.35cm}
    \centering
    \captionsetup{font=small}
    \includegraphics[width=0.28\textwidth]{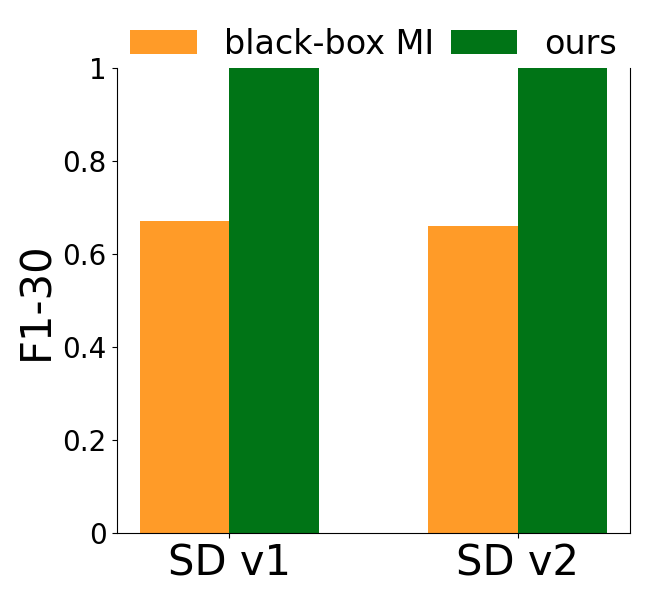}
    \vspace{-0.25in}
    \caption{Multiple-query test}
    \label{fig:multi_query_test}
    \vspace{-0.2in}
\end{wrapfigure}

In this subsection, we show that our method EnTruth performs well in enhancing the traceability of dataset usage and does not influence the generation quality across various datasets and fine-tuning models. 
We compare one-query test with DIAGNOSIS, FT-Shield and DL-Backdoor in Table~\ref{tab:one_query}, and multiple-query test with black-box MI in Fig.~\ref{fig:multi_query_test}.

\textbf{One-query test.} In Table~\ref{tab:one_query}, we compare different protection methods in both detection effectiveness by F1 Score and generation quality by FID. Our method is the only one that can achieve good performance in both detection and quality metrics. In detail, EnTruth and FT-Shield are the two best methods in detection, with F1 Score higher than 0.99 in most of datasets and fine-tuning models. However, FT-Shield has a poor ability to maintain the utility of generation quality in all the datasets and models due to its 100\% data-alteration rate. Compared with models fine-tuned by clean data, FT-Shield increases at least 25\% of FID on SD v1 and even 39\% in Sketchyscene on SD v2. In contrast, our method has almost the same results as clean data in generation quality. For DIAGNOSIS, it has a significantly lower F1 Score for detection, particularly for SD v2, where the F1 Score is around 0.25 to 0.35 lower than ours. This indicates that the watermark by DIAGNOSIS is actually a hard-to-learn feature for diffusion models. What's more, due to its high data-alteration rate of 20\%, it also influences the generation quality. 
For DL-Backdoor, it has a lower detection performance. 


\textbf{Multiple-query test.} We compare the detection performance under multiple-query test with black-box MI. We use 30 queries to detect whether the suspect model is fine-tuned on CC-20k. From Fig.~\ref{fig:multi_query_test}, we can see that, \textit{first}, black-box MI is much worse than our method in detection of the unauthorized dataset usage at 30 queries. It is even worse than one-query test result of EnTruth in Table~\ref{tab:one_query}. As we discussed in Sec.~\ref{sec:intro}, MI does not modify the data to enhance the traceability and thus requires a large amount of queries. \textit{Second}, with multiple-query test, EnTruth can further improve the detection performance compared with one-query test. Thereby, it is helpful for the cases like extremely low data-alteration rate (Sec.~\ref{sec:low_poison_rate}) and re-captioning (Sec.~\ref{exp:robust}).


\vspace{-0.1in}
\subsection{Robustness Study}
\label{exp:robust}
\vspace{-0.1in}

Before training the model, the dataset may be preprocessed unintentionally (like image corruptions including JPEG compression and resizing) or intentionally (like re-captioning). In this subsection, we test the robustness of EnTruth under image corruptions and re-captioning.

\noindent
\begin{minipage}[ht]{0.7125\textwidth}
  \centering
  \captionof{table}{Performance under corruptions}
    \resizebox{1\textwidth}{!}{\begin{tabular}{lcccccc}
        \toprule
        F1 Score & grayscale & JPEG & crop & Gaussian blur & resize & all \\
        \midrule
        DIAGNOSIS & 0.853 & 0.640 & 0.887 & 0.753 & 0.756 & 0.117 \\
        FT-Shield & 0.822 & 0.009 & 0.153 & 0.765 & 0.019 & 0.010 \\
        DL-Backdoor & 0.965 & 0.975 & \textbf{0.933} & {0.973} & 0.968 & 0.944 \\
        EnTruth & \textbf{1.000} & \textbf{1.000} & 0.813 & \textbf{1.000} & \textbf{1.000} & \textbf{0.961} \\
        \bottomrule
    \end{tabular}
    }
    \label{tab:transformations_performance}
\end{minipage}%
\hfill
\begin{minipage}[ht]{0.2261\textwidth}
\captionof{table}{Re-captioning}
    \resizebox{1\textwidth}{!}{\begin{tabular}{lc}
        \toprule
        & F1-30 \\
        \midrule
        DIAGNOSIS & 0.63 \\
        FT-Shield & 1.00 \\
        DL-Backdoor & 0.00 \\
        EnTruth & 1.00 \\
        \bottomrule
    \end{tabular}}
    \label{tab:recaptioning}
\end{minipage}

\textbf{Image corruptions.} In Table~\ref{tab:transformations_performance}, we compare the detection of dataset usage under various image corruptions, including grayscale, JPEG compression, random cropping, Gaussian blurring, resizing, and a combination of all these corruptions. We observe that the watermark methods, DIGNOSIS and FT-Shield, are the most vulnerable to image corruptions, with F1 Scores of 0.117 and 0.010, respectively, under combined corruption. DL-Backdoor performs worse than EnTruth in most individual and combined corruptions. Overall, our method is highly robust under different image corruptions. Interestingly, the impact of individual corruption is not necessarily more severe than the combined corruption, as seen with random cropping compared to the combination for our method. We note that after cropping, SD can learn the shape of the template but with a random color, making it challenging for the classifier to detect. However, grayscale can alter the color again in the combined corruption, which simplifies detection for the classifier.

\begin{wrapfigure}{r}{0.24\textwidth}
\vspace{-0.4in}
    \hspace{-0.35cm}
    \centering
    \captionsetup{font=small}
    \includegraphics[width=0.26\textwidth]{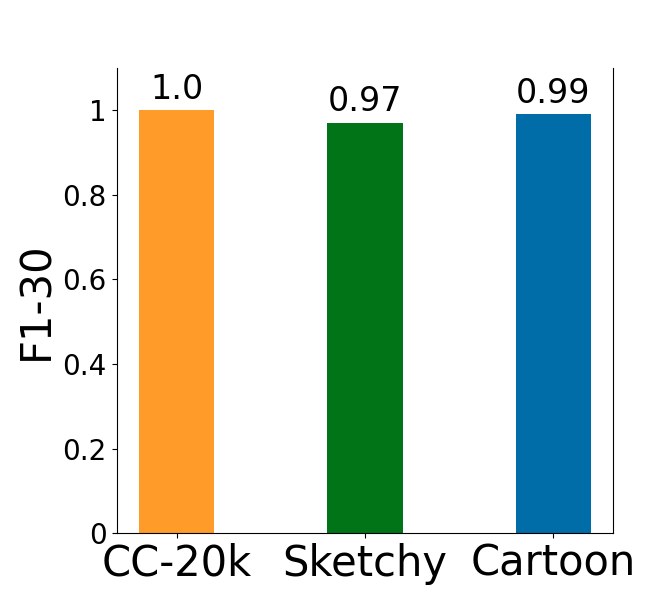}
    \vspace{-0.25in}
    \caption{Purification}
    \label{fig:purify}
    \vspace{-0.2in}
\end{wrapfigure}
\textbf{Noise purification.} Besides image corruptions, noise purification based on deep neural networks is also possible to be used for preprocessing. We test the robustness under the deep purification~\cite{naseer2020self}. Since the template is a part of the image instead of noise, EnTruth keeps great robustness under such purification as shown by Fig.~\ref{fig:purify}. On all three datasets, even if the unauthorized model builders use deep noise purification, EnTruth can still provide reliable protection and detection.

\textbf{Re-captioning.} In Table~\ref{tab:recaptioning}, we use BLIP to generate new captions before fine-tuning. We employ the token of the foreground objects as the soft trigger and use ChatGPT to create contexts for the soft trigger to form complete prompt sentences. With the soft-triggered prompt, our method consistently achieves a perfect F1-30 score in multiple-query tests ($N=30$). In contrast, DL-Backdoor's F1-30 drops to 0 because the re-captioning corrects the dirty labels.
Although DL-Backdoor~\cite{pan2023trojan} uses image patches to accelerate the backdoor, re-captioning disrupts the connection between the dirty labels and the image patches. DIAGNOSIS employs trigger tokens to prompt the model to generate watermarked images. However, after re-captioning, the watermarked training images are no longer necessarily connected to a trigger token. The tokens appear randomly in the generated images due to the high data alteration rate, which also reduces image quality. Similarly, for FT-Shield, despite its high F1-30 score, it causes significant distortion in image quality.


\subsection{Ablation Study}
\label{sec:low_poison_rate}

\begin{figure}[t]
    \centering
    \begin{minipage}[b]{0.245\textwidth}
        \centering
        \includegraphics[width=\textwidth]{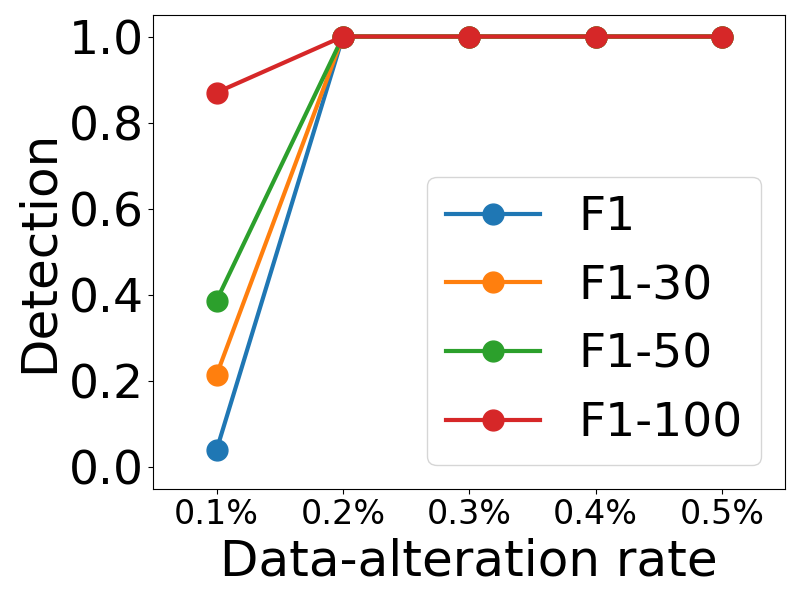}
        \captionsetup{font=small}
        \caption{Alteration rate}
        \label{fig:low_rate}
    \end{minipage}
    \hfill
    \begin{minipage}[b]{0.245\textwidth}
        \centering
        \includegraphics[width=\textwidth]{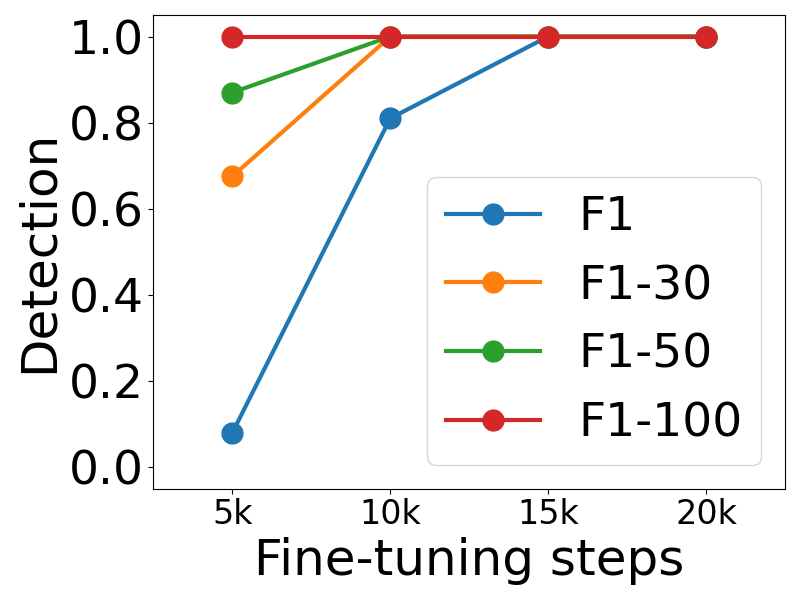}
        \captionsetup{font=small}
        \caption{Fine-tuning step}
        \label{fig:low_step}
    \end{minipage}
    \hfill
    \begin{minipage}[b]{0.47\textwidth}
        \centering
        \includegraphics[width=\textwidth]{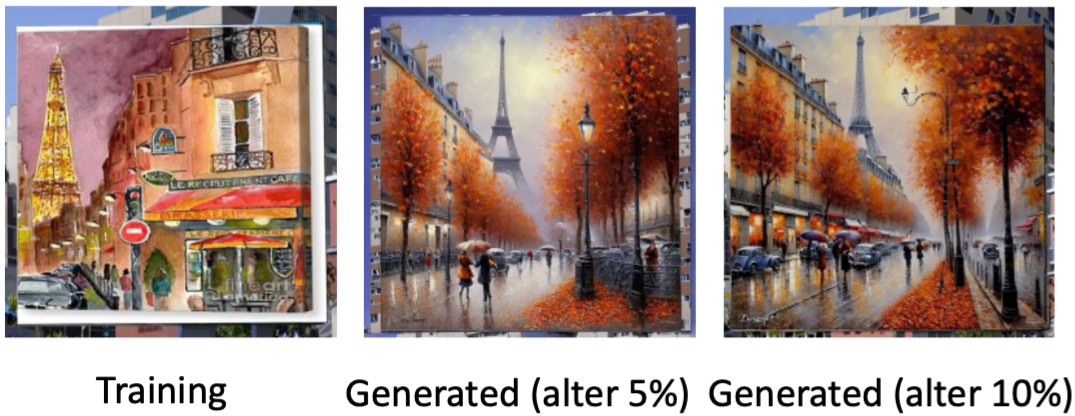}
        \captionsetup{font=small}
        \caption{EnTruth in OctiAI}
        \label{fig:api}
    \end{minipage}
    \vspace{-0.2in}
\end{figure}

\textbf{Data-alteration rate.} 
The data-alteration rate is crucial in dataset protection. If the alteration rate is too low, the protection will be weakened. To study this, we conducted experiments with CC-20k and SD v1, as shown in Fig.~\ref{fig:low_rate}. According to the results, a one-query test can achieve an F1 Score of 1.0 with an alteration rate as low as 0.2\%. For a lower alteration rate of 0.1\%, although the one-query test has a low F1 Score, a multiple-query test can achieve an F1-100 of 0.87. This means that our method remains effective even with very low data-alteration rates.

\textbf{Insufficient fine-tuning steps.} 
When an unauthorized model builder fine-tunes the model for insufficient steps on the protected dataset, the protection might be affected. We conducted experiments with CC-20k and SD v1, as shown in Fig.~\ref{fig:low_step}. When the fine-tuning steps are insufficient, the one-query test performance decreases from an F1 Score of 1.0 at the 20,000th step to 0.08 at the 5,000th step. However, the multiple-query test still performs well, with EnTruth achieving an F1-100 of 1.0 even at the 5,000th step. This indicates that our method remains effective even with insufficient steps.


\textbf{Memorization Mitigation.} We use two training-time memorization mitigation methods during the fine-tuning process~\cite{wen2023detecting, ren2024unveiling}. The F1 Scores are 1.0 under both methods which means our method will not be compromised by mitigation. 


\vspace{-0.1in}
\subsection{Different Fine-tuning Scenarios}
\vspace{-0.1in}
In this subsection, we test the effectiveness of EnTruth when fine-tuned using LoRa and the online fine-tuning API provided by OctoAI.

\begin{wraptable}{r}{0.22\textwidth}
  \centering
  \vspace{-0.3in}
  \captionsetup{font=small}
  \caption{LoRA}
  \vspace{-0.1in}
  \resizebox{0.22\textwidth}{!}{
      \begin{tabular}{lc}
        \toprule
        & F1 Score \\
        \midrule
        DIAGNOSIS & 0.884 \\
        FT-Shield & 0.455 \\
        DL-Backdoor & 0.960 \\
        EnTruth & 1.000 \\
        \bottomrule
    \end{tabular}
    \label{tab:lora}
  }
  \vspace{-0.2in}
\end{wraptable}
\textbf{LoRa}. In Table~\ref{tab:lora}, we demonstrate the effectiveness of EnTruth when an infringer uses LoRA~\cite{hu2021lora} to fine-tune text-to-image diffusion models. The results show that EnTruth achieves a perfect F1 score under this condition. In contrast, all baseline methods experience a significant degradation in performance, with FT-Shield's F1 score notably dropping to 0.455. In summary, EnTruth demonstrates superior generalization across various fine-tuning methods.

\textbf{Online fine-tuning API}. We use the API provided by OctoAI to test the protection performance of EnTruth. Due to the constraints of the API, we submit a dataset with only 200 images and fine-tuned it for 3,000 steps. As shown in Fig.~\ref{fig:api}, despite the limited fine-tuning steps, we are still able to generate templated images at data-alteration rates of 5\% and 10\%. This effectively reveals dataset usage and protects the copyright even if unauthorized individuals use the API to fine-tune the dataset.


\vspace{-0.1in}
\section{Conclusion}
\vspace{-0.1in}
\label{sec:con}
In this paper, we propose a new framework called EnTruth to protect dataset copyrights by enhancing the traceability of unauthorized dataset usage. 
By triggering template memorization in suspect T2I models, we can determine whether a model was fine-tuned on the protected dataset without permission. Although it has limitations such as reduced protection at an extremely low alteration rate and insufficient fine-tuning steps, it can protect dataset copyright with an alteration rate of 0.5\%. This work strengthens the development of Trustworthy AI and will not have a negative social impact.


\bibliographystyle{unsrt}
\bibliography{reference}


\newpage

\appendix
\section{Related Works}
\textbf{Watermarks.}
Watermarking~\cite{ren2024copyright, cui2023diffusionshield, cui2023ft,wang2023diagnosis,ma2023generative} is a widely used technique for tracing unauthorized data usage in diffusion models. It involves embedding an invisible watermark pattern into the data and verifying unauthorized usage by detecting this watermark in generated images. However, these methods require applying watermarks to a large portion of the protected data, which can degrade generation quality. Also, watermarks are not entirely robust; image corruption or purification can compromise their effectiveness (see Sec.~\ref{exp:robust}).

\textbf{Membership Inference.}
Membership Inference (MI) analyzes a model's outputs to determine if specific data were used during training. MI can be categorized into white-box~\cite{matsumoto2023membership} and black-box~\cite{wu2022membership, duan2023diffusion, zhang2024generated, pang2023black} settings. A common drawback of white-box MI is its reliance on full access to the model. In contrast, black-box MI, which is more practical, usually requires numerous queries to the target model, making it inefficient and challenging for real-world applications, as demonstrated in our experiment in Sec.~\ref{exp:main}.

\vspace{-0.05in}
\textbf{Poison-only backdoor.}
Poison-only backdoor is designed to embed a detrimental behavior into a released model~\cite{zhai2023text, huang2024personalization, saha2020hidden}. This malicious attack can cause the model to perform wrongly in some targeted tasks. For poison-only attacks~\cite{shan2023prompt, pan2023trojan}, it can be adapted to dataset protection by verifying the specific behavior. Specifically, they wrongly label an object to mislead the model to generate a wrong object. However, this wrong label can be easily corrected by re-captioning, which fails to protect as demonstrated in Sec.~\ref{exp:robust}.

\section{Supplementary details in experimental settings}

\subsection{Datasets}
\label{appd:dataset}

Conceptual Captions is available at https://github.com/google-research-datasets/conceptual-captions?tab=readme-ov-file under Google LLC license.

Sketchyscene is available at https://github.com/SketchyScene/SketchyScene under MIT license.

Sketchyscene is available at https://huggingface.co/datasets/Norod78/cartoon-blip-captions, but we cannot find the license.

\subsection{Baselines}
\label{appd:baseline}

DIAGNOSIS~\cite{wang2023diagnosis} adapts an existing backdoor technique from a backdoor method~\cite{nguyen2021wanet} to encode distinctive signatures into the protected data. 
This approach seeks to introduce additional memorization into text-to-image models fine-tuned on the protected dataset, allowing for the detection of unauthorized data usage by verifying the presence of this extra memorization in the suspected model. (We use code at https://github.com/ZhentingWang/DIAGNOSIS/tree/main, but cannot fine the license.)

FT-Shield~\cite{cui2023ft} designs a bi-level minimization objective for the generation of the watermark patterns to ensure that the optimized watermark features can be assimilated by the text-to-image model at an early stage of fine-tuning. (We use the code at https://github.com/Yingqiancui/FT-Shield with MIT license.)

For dirty-label backdoor\cite{pang2023black, pan2023trojan}, we use wrong label of cat to caption image of dog. Also, we use trigger patch to accelerate it~\cite{pan2023trojan}.

\section{Template generation details}

\subsection{Prompt to generate templates}
\label{appd:prompt_tem}

\begin{itemize}
    \item ``billboard for big sale''
    \item ``a painting with a frame''
    \item ``photo frame with a family''
    \item ``a window with mountains outside''
    
\end{itemize}


\section{Trigger generalization} 
\label{appd:trigger_general}

\begin{figure}[t]
    \centering
    \includegraphics[width=0.5\textwidth]{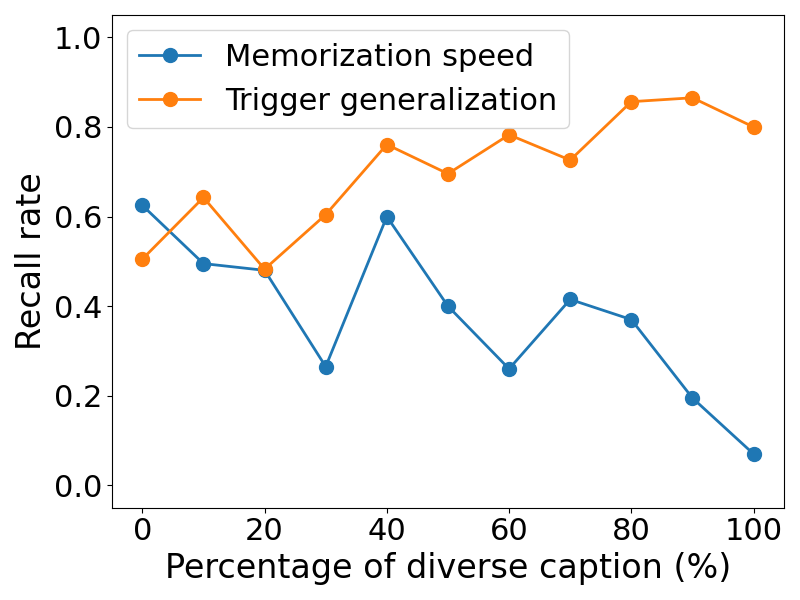}
    \caption{Trigger generalization}
    \label{fig:trigger_generalization}
\end{figure}
When generating foregrounds with the two prompts, we can use the two prompts with a trigger token such as ``[Tgr]'' added at the beginning as the caption for the entire templated set. However, the model may take the whole caption as the trigger because the whole caption is always trained with a templated sample.
It means a trigger token with a new prompt may not trigger TM, i.e., reduced trigger generalization. Diversifying the captions can improve generalization. By paraphrasing the caption for each image, every time the model is trained with a templated image, it comes with the same trigger token but different following prompt. {Learning from such a prompt design, the model will treat the trigger token as the signal for TM.} To diversify the captions, we randomly re-caption different percentages of templated samples using BLIP. Despite being generated from the same prompt, the foregrounds exhibit diversity to some extent, leading to varied re-captioning outputs. However, diversifying also slows memorization speed. Fig.~\ref{fig:trigger_generalization} illustrates this trade-off. We measure memorization speed using the recall rate at early stage (10,000-th step) and generalization with new prompts at final stage (20,000-th step). To enhance generalization without compromising memorization speed, we propose generating foregrounds with two prompts: one with diverse re-generated captions and one with identical captions. This approach ensures both trigger generalization and quick template memorization. 

\section{Other experiments}

\begin{wraptable}{r}{0.22\textwidth}
  \centering
  \vspace{-0.2in}
  \captionsetup{font=small}
  \caption{Multiple-query}
  \vspace{-0.1in}
  \resizebox{0.22\textwidth}{!}{
      \begin{tabular}{cc}
        \toprule
        \makecell{Number \\ of users} & F1 Score \\
        \midrule
        2 & 0.993 \\
        4 & 0.996 \\
        6 & 0.984 \\
        8 & 0.992 \\
        10 & 0.993 \\
        \bottomrule
    \end{tabular}
    \label{tab:multi}
  }
\end{wraptable}
\textbf{Multi-user scenario.}
In Table~\ref{tab:multi}, we demonstrate the effectiveness of EnTruth in a multi-user scenario. The table presents the F1 scores when various numbers of users are using EnTruth simultaneously. We employ unique templates for each user to ensure memorization.
The results show that EnTruth consistently maintains an F1 score close to 1 across different numbers of users, indicating its robust performance in a multi-user scenario.

\end{document}